\documentclass{article}

\PassOptionsToPackage{compress}{natbib}

\usepackage[preprint]{neurips_data_2023}

\usepackage[utf8]{inputenc} 
\usepackage[T1]{fontenc}    
\usepackage{hyperref}       
\usepackage{url}            
\usepackage{booktabs}       
\usepackage{amsfonts}       
\usepackage{nicefrac}       
\usepackage{microtype}      
\usepackage{xcolor}         
\usepackage{graphicx}
\usepackage{amsmath}
\usepackage{enumitem}
\usepackage{caption}
\usepackage{pifont}
\def\sensorium{\texttt{SENSORIUM}}
\usepackage{tikz}
    \tikzset{%
        baseline,
        inner sep=2pt,
        minimum height=12pt,
        rounded corners=2pt  
    }

\hypersetup{
    colorlinks,
    linkcolor={red!50!black},
    citecolor={blue!50!black},
    urlcolor={blue!80!black}
}

\newcommand{\fig}[1]{Fig.~\ref{#1}}
\newcommand{\sect}[1]{Section~\ref{#1}}
\newcommand{\tabl}[1]{Table~\ref{#1}}

\title{Retrospective for the Dynamic Sensorium Competition for
predicting large-scale mouse primary visual cortex
activity from videos}
\begin{document}
\maketitle

{\bf\vspace{-44pt}
\begin{center}    
Polina Turishcheva\textsuperscript{*,1,\ding{41}}, 
Paul G. Fahey\textsuperscript{*,2--5,\ding{41}}, 
Michaela Vystr\v{c}ilová\textsuperscript{1}, 
Laura Hansel\textsuperscript{1}, 
\mbox{Rachel Froebe\textsuperscript{2--5}},  
Kayla Ponder\textsuperscript{2}, 
Yongrong Qiu\textsuperscript{1,3--5}, 
Konstantin F. Willeke\textsuperscript{1,6,7}, 
\mbox{Mohammad Bashiri\textsuperscript{1,6,7}} 
Ruslan Baikulov\textsuperscript{8}, 
Yu Zhu\textsuperscript{9,10}, 
Lei Ma\textsuperscript{10}, 
Shan Yu\textsuperscript{9}, 
Tiejun Huang\textsuperscript{10}, 
Bryan M. Li\textsuperscript{11, 12}, 
Wolf De Wulf\textsuperscript{12}, 
Nina Kudryashova\textsuperscript{12}, 
Matthias H. Hennig\textsuperscript{12}, 
\mbox{Nathalie L. Rochefort\textsuperscript{13,14}}, 
Arno Onken\textsuperscript{12}, 
Eric Wang\textsuperscript{2}, 
Zhiwei Ding\textsuperscript{2}, 
\mbox{Andreas S. Tolias\textsuperscript{2--5,15}}, 
Fabian H. Sinz\textsuperscript{1,2,6,7}, 
Alexander S Ecker\textsuperscript{1,16,\ding{41}} 
\end{center}
}  
{\footnotesize
\textsuperscript{\bf 1}\,Institute of Computer Science and Campus Institute Data Science, University of Göttingen, Germany, 
\textsuperscript{\bf 2}\,Department of Neuroscience \& Center for Neuroscience and Artificial Intelligence, Baylor College of Medicine, Houston, Texas, USA,
\textsuperscript{\bf 3}\,Department of Ophthalmology, Byers Eye Institute, Stanford University School of Medicine, Stanford, CA, US,
\textsuperscript{\bf 4}\,Stanford Bio-X, Stanford University, Stanford, CA, US,
\textsuperscript{\bf 5}\,Wu Tsai Neurosciences Institute, Stanford University, Stanford, CA, US,
\textsuperscript{\bf 6}\,International Max Planck Research School for Intelligent Systems, Tübingen, Germany
\textsuperscript{\bf 7}\,Institute for Bioinformatics and Medical Informatics, Tübingen University, Germany, 
\textsuperscript{\bf 8}\,lRomul, Russia,
\textsuperscript{\bf 9}\,Institute of Automation, Chinese Academy of Sciences, China
\textsuperscript{\bf 10}\,Beijing Academy of Artificial Intelligence, China
\textsuperscript{\bf 11}\,The Alan Turing Institute, UK
\textsuperscript{\bf 12}\,School of Informatics, University of Edinburgh, UK
\textsuperscript{\bf 13}\,Centre for Discovery Brain Sciences, University of Edinburgh, UK
\textsuperscript{\bf 14}\,Simons Initiative for the Developing Brain, University of Edinburgh, UK
\textsuperscript{\bf 15}\,Department of Electrical Engineering, Stanford University, Stanford, CA, US,
\textsuperscript{\bf 16}\,Max Planck Institute for Dynamics and Self-Organization, Göttingen, Germany. 
\textsuperscript{\bf *}\,Equal contribution.
\textsuperscript{\ding{41}}\,\href{mailto:turishcheva@cs.uni-goettingen.de}{turishcheva@cs.uni-goettingen.de},
\href{mailto:pgfahey@stanford.edu}{pgfahey@stanford.edu}, \href{mailto:ecker@cs.uni-goettingen.de}{ecker@cs.uni-goettingen.de}
}
\vspace{5pt}
\begin{abstract}
Understanding how biological visual systems process information is challenging because of the nonlinear relationship between visual input and neuronal responses. 
Artificial neural networks allow computational neuroscientists to create predictive models that connect biological and machine vision.
Machine learning has benefited tremendously from benchmarks that compare different model on the same task under standardized conditions. 
However, there was no standardized benchmark to identify state-of-the-art dynamic models of the mouse visual system.
To address this gap, we established the \sensorium~ 2023 Benchmark Competition with dynamic input, featuring a new large-scale dataset from the primary visual cortex of ten mice. 
This dataset includes responses from 78,853 neurons to 2 hours of dynamic stimuli per neuron, together with the behavioral measurements such as running speed, pupil dilation, and eye movements.
The competition ranked models in two tracks based on predictive performance for neuronal responses on a held-out test set: one focusing on predicting in-domain natural stimuli and another on out-of-distribution (OOD) stimuli to assess model generalization.
As part of the NeurIPS 2023 competition track, we received more than 160 model submissions from 22 teams. 
Several new architectures for predictive models were proposed, and the winning teams improved the previous state-of-the-art model by 50\%. 
Access to the dataset as well as the benchmarking infrastructure will remain online at \url{www.sensorium-competition.net}.
\end{abstract}


\section{Introduction}
\label{intro}

Understanding visual system processing is a longstanding goal in neuroscience.
One way to approach the problem are neural system identification approaches which make predictions of neuronal activity from stimuli or other sources quantitative and testable~\citep[reviewed in][]{wu2006complete}. 
Various system identification methods have been used in systems neuroscience, including linear-nonlinear (LN) models \citep{simoncelli2004characterization, Jones1987-sn,Heeger1992-ig, Heeger1992-xx}, energy models \citep{Adelson1985-re}, subunit models \citep{liu2017inference, rust2005spatiotemporal, touryan2005spatial, vintch2015convolutional}, Bayesian models \citep{walker2020neural, george2005hierarchical,wu2023bayesian, bashiri2021flow}, redundancy reduction models \citep{perrone2015redundancy}, and predictive coding models \citep{marques2018functional}.
In recent years, deep learning models, especially convolutional neural networks (CNNs) trained on image recognition tasks \citep{Yamins2014,Cadieu2014,Cadena2019,pogoncheff2023explaining} or predicting neural responses \citep{Cadena2019, Antolik2016,batty2017multilayer,McIntosh2016,Klindt2017,Kindel2017,burg2021learning, Lurz2020-ua, bashiri2021flow,Zhang2018-cs,Cowley2020neurips,Ecker2018, sinz2018stimulus, Walker2019-oq,Franke2022, wang2023towards, Fu2023.03.13.532473, ding2023bipartite}, have significantly advanced predictive model performance. 
More recently, transformer-based architectures have emerged as a promising alternative \citep{li2023vt,azabou2024unified,antoniades2023neuroformer}.



In machine learning and beyond, standardized large-scale benchmarks foster continuous improvements in predictive models through fair and competitive comparisons \citep{dean2018new}. 
Within the realm of computational neuroscience, several benchmarks have been established recently. 
An early effort was the Berkeley Neural Prediction Challenge\footnote{\url{https://neuralprediction.org/npc/con.php}},
which provided public training data and secret test set responses to evaluate models of neurons from primary visual cortex, primary auditory cortex and field L in the songbird brain. 
More recent efforts include Brain-Score \citep{Schrimpf2018, Schrimpf2020-hd}, Neural Latents '21 \citep{pei2021neural}, Algonauts \citep{Cichy2019-re, Cichy2021-lr, gifford2023algonauts} and \sensorium~2022 \citep{willeke2022sensorium}. 
However, with the exception of the Berkeley Neural Prediction Challenge, which is limited to 12 cells, no public benchmark existed that focused on predicting single neuron responses in the early visual system to video (spatio-temporal) stimuli. 

Since we all live in a non-static world, dynamic stimuli are more relevant and our models should be able to predict neural responses over time in response to these time-varying inputs \citep{sinz2018stimulus, wang2023towards, batty2017multilayer, McIntosh2016, zheng, qiu2023efficient, hoefling2022chromatic, vystrcilova2024}.
Even though recent high-throughput recording techniques have led to the release of large datasets like the MICrONS calcium imaging dataset \citep{microns2021functional} and Neuropixel datasets from the Allen Brain Observatory \citep{de2020large, Siegle2021-en}, the absence of a withheld test set and the corresponding benchmark infrastructure hinders a fair comparison between different models.

To fill this gap, we established the \sensorium~ 2023 competition, with the goal to compare large-scale models predicting single-neuron responses to dynamic stimuli.
The NeurIPS 2023 competition received over 160 model submissions from 22 teams and resulted in new state-of-the-art predictive models that improved over the competition baseline by 50\%.
Moreover, these models also led to a 70\% improved predictions on out-of-domain stimuli, suggesting that more predictive models on natural scenes also generalize better to other stimuli.



\begin{figure*}[t]
    \centering
    \includegraphics[trim=0 0 0 0, clip,width=\linewidth]{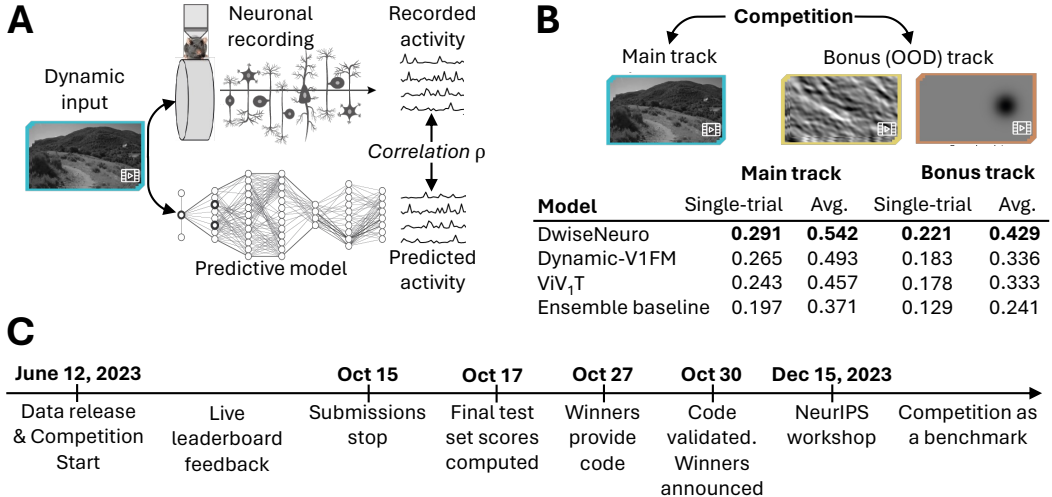}
    \caption{\textbf{A schematic illustration of the \sensorium~ competition.}
        \textbf{A:} Competition idea.
         We provide large-scale datasets of neuronal activity in the mice visual cortex in response to natural videos. 
         The competition participants were tasked to find the best models to predict neuronal activity for a set of videos for which we withheld the ground truth.
         \textbf{B:}  Tracks and leaderboard.
         \textbf{C:} Timeline.
    }\label{fig:competition_overview}
\end{figure*}

\section{Sensorium Competition Overview}

The goal of the competition was to advance models that predict neuronal responses of several thousand neurons in mouse primary visual cortex to natural and artificially generated movies.
We collected and released a comprehensive dataset consisting of visual stimuli and corresponding neuronal responses for training (Section~\ref{sec:data}). 
This dataset included a dedicated test set, for which we released only the visual stimuli but withheld the neuronal responses to be able to compare models in a fair way (\fig{fig:competition_overview}A). 
To assess model performance, participants submitted their predictions on the test set to our online benchmark website for evaluation and ranking against other submissions.\footnote{Our benchmark webpage is based on Codalab Competitions \citep{pavao2022codalab} available under the Apache License 2.0 \url{https://github.com/codalab/codalab-competitions}} 
The test set consisted of two parts: a \emph{live test set} and a \emph{final test set}. 
The live test set was used during the competition to give feedback to the participants on their model's performance via a leaderboard. 
The final test set was used only after the end of submissions to determine the final winners (\fig{fig:competition_overview}C).
From each team only the best-performing submission on the live test set was evaluated on the final test set. 
To facilitate participation from both neuroscientists and machine learning practitioners, we developed user-friendly APIs that streamline data loading, model training, and submission.\footnote{\url{https://github.com/ecker-lab/sensorium_2023}}

The competition consisted of two tracks: The \emph{main track} and the \emph{bonus track} (\fig{fig:competition_overview}B).
The main track entailed predicting responses on natural movie stimuli, the same type of stimuli available for model training, but different movie instances. 
The bonus track required predicting out-of-distribution (OOD) stimuli for which no ground truth responses of the neurons were provided in the training set. 
This bonus track tests a model's ability to generalize beyond the training data. 

The competition ran from June 12 to October 15, 2023, culminating in a NeurIPS 2023 conference workshop where the winning teams presented their approaches and insights.
The benchmark platform will continue to track advancements in developing models for the mouse primary visual cortex. 
In the following, we describe the dataset (\sect{sec:data}) and evaluation metrics (\sect{metrics}), the baseline (\sect{baseline}) and winning models (\sect{results}) and report on the results and learnings (\sect{reflections}).

\section{Dataset}
\label{sec:data}
\label{data}
\begin{figure*}[t]
    \begin{minipage}[t]{1\linewidth}
        \vspace{0 pt}
        \centering
        \vspace{0 pt}
        \centering
        \includegraphics[trim=0 0 0 0, clip,width=\linewidth]{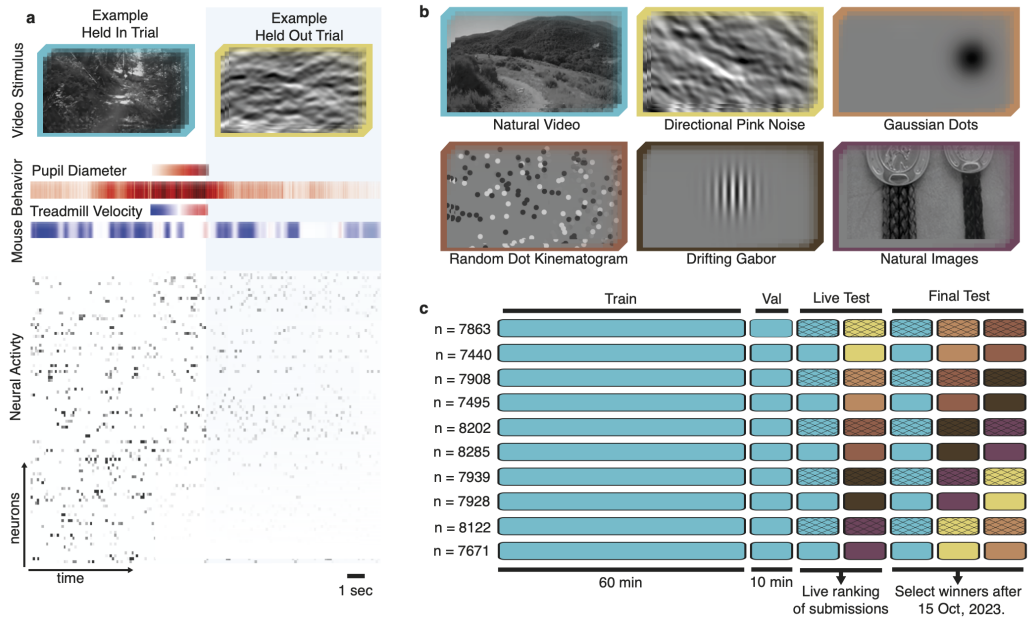}
    \end{minipage}\hfill
    \begin{minipage}[t]{\linewidth}
        \vspace{5 pt}
        \caption{\textbf{Overview of the data.}
             \textbf{a}, Example stimulus frames, behavior (pupil position not depicted) and neural activity. \textbf{b}, Representative frames from natural video and five OOD stimuli. \textbf{c}, Stimulus composition (color) and availability for all five scans in ten animals.  $n$ is number of neurons per scan. 
             The crossed elements were used for live and final test sets in the competition evaluation.}
   \label{fig:data_overview}
    \end{minipage}
\end{figure*}
We recorded\footnote{Full neuroscience methods are available at \cite{turishcheva2023dynamic}} neuronal activity in response to natural movie stimuli as well as several behavioral variables, which are commonly used as a proxy of modulatory effects of neuronal responses \citep{niell2010modulation, reimer2014pupil}. 
In general terms, neural predictive models capture neural responses $\mathbf{r} \in \mathbb{R}^{n\times t}$ of $n$ neurons for $t$ timepoints as a function $\mathbf{f}_\theta(\mathbf{x}, \mathbf{b})$ of both natural movie stimuli $\mathbf{x} \in \mathbb{R}^{w\times h\times t}$, where $w$ and $h$ are video width and height, and behavioral variables $\mathbf{b} \in \mathbb{R}^{k\times t}$, where $k=4$ is the number of behavioral variables (see below).  

\textbf{Movie stimuli.}  
We sampled natural dynamic stimuli from cinematic movies and the Sports-1M dataset \citep{Karpathy2014-gu}, as described by \citet{microns2021functional}.
Following earlier work \citep{wang2023towards}, we showed five additional stimulus types for the bonus track (\fig{fig:data_overview}b): directional pink noise \citep{microns2021functional}, flashing Gaussian dots, random dot kinematograms \citep{Morrone2000-ov}, drifting Gabors \citep{Petkov2007-kv}, and natural images from ImageNet \citep{Russakovsky2015-xi, Walker2019-oq}.
Stimuli were converted to grayscale and presented to mice in clips lasting $\sim8$ to $11$ seconds, at 30 frames per second.

\textbf{Neuronal responses.} 
Using a wide-field two-photon microscope \citep{sofroniew2016large}, we recorded the responses of excitatory neurons at 8\,Hz in layers 2--5 of the right primary visual cortex in awake, head-fixed, behaving mice using calcium imaging. Neuronal activity was extracted as described previously \citep{wang2023towards} and upsampled to 30\,Hz to be at the same frame rate as the visual stimuli. 
We also released the anatomical coordinates of the recorded neurons. 

\textbf{Behavioral variables.} 
We measured four behavioral variables: \emph{locomotion speed}, recorded from a cylindrical treadmill at 100\,Hz and resampled to 30\,Hz, and \emph{pupil size, horizontal and vertical pupil center position}, each extracted at 20\,Hz from video recordings of the eye and resampled to 30 Hz. 

\textbf{Datasets and splits.}
Our complete dataset consists of ten recordings from ten different animals, in total containing the activity of 78,853 neurons to a total of $\sim$1200 minutes of dynamic stimuli, with $\sim$120 minutes per recording.
The recordings were collected and released explicitly for this competition. None of them had been published before. Each recording had four components (\fig{fig:data_overview}c): 

\begin{itemize}[leftmargin=15ex]
    \item[\textbf{Training set:}] 60 minutes of natural movies, one repeat each (60 minutes total).
    \item[\textbf{Validation set:}] 1 minute of natural movies, ten repeats each (10 minutes total). 
    \item[\textbf{Live test set:}] 1 minute of natural movies and 1 minute of OOD stimuli, ten repeats each (20 minutes total).  Each OOD stimulus type is presented only in one of the five recordings.
    \item[\textbf{Final test set:}] 1 minute of natural movies and 2 minutes of OOD stimuli, ten repeats each (30 minutes total).  Each OOD stimulus type is presented in two of the five recordings. 
\end{itemize}
For the training set and validation set, the stimulus frames, neuronal responses, and behavioral variables are released for model training and evaluation by the participants, and are not included in the competition performance metrics.  
For the five mice included in the competition evaluation, the train and validation sets contain only natural movies but not the OOD stimuli. 
For the other five mice, all stimuli and responses, including test sets and OOD stimuli, were released.

\textbf{Code and data availability.}
The competition website and data are available at \url{https://www.sensorium-competition.net/}. 
Starter kit and benchmark code are available at \url{https://github.com/ecker-lab/sensorium_2023}.

\section{Competition evaluation}
\label{metrics}
Similar to \sensorium\ 2022, we used the correlation coefficient between predicted and measured responses to evaluate the models. 
Since it is bounded between $-1$ and $1$, the correlation coefficient is straightforward to interpret. 
Because neuronal responses fluctuate from trial to trial, the correlation between model predictions and single-trial neuronal responses typically do not reach the upper bound of 1 even for a perfect model. 
This trial-to-trial variability can be reduced by averaging over repeated presentations of the same stimulus. 
However, in this case, also the contributions from behavioral states are reduced since these cannot be repeated easily during uncontrolled behavior. 
We therefore computed two metrics: \emph{single-trial correlation} and \emph{correlation to average}.


\textbf{Single-trial correlation}, $\rho_{\textrm{st}}$, on the natural video final test set was used to determine competition winners for the main track. 
We also computed the single-trial correlation metric for each of the five OOD stimulus types in the test sets separately. 
The average single-trial correlation across all five final OOD test sets were used to determine the competition winner for the bonus track.
Single trial correlation is sensitive to variation between individual trials and computes correlation between single-trial model output (prediction) $o_{ij}$ and single-trial neuronal responses $r_{ij}$, as
\begin{equation}
    \rho_{\textrm{st}} = \textrm{corr}(\mathbf{r}_{\textrm{st}}, \mathbf{o}_{\textrm{st}}) = \frac{\sum_{i,j}(r_{ij} - \bar{r})(o_{ij} - \bar{o})}{\sqrt{\sum_{i,j}(r_{ij} - \bar{r})^2 \sum_{i,j}(o_{ij} - \bar{o})^2}},
\end{equation}

where $r_{ij}$ is the $i$-th frame of $j$-th video repeat, $o_{ij}$ is the corresponding prediction, which can vary between stimulus repeats as the behavioral variables are not controlled. 
The variable $\bar{r}$ is the average response to all the videos in the corresponding test subset across all stimuli and repeats, and $\bar{o}$ is the average prediction for the same videos and repeats
.
The single-trial correlation $\rho_{\textrm{st}}$ was computed independently per neuron and then averaged across all neurons to produce the final metric. 

\textbf{Correlation to average}, $\rho_{\textrm{ta}}$, provides a more interpretable metric by accounting for trial-to-trial variability through averaging neuronal responses over repeated presentations of the same stimulus. 
As a result, a perfect model would have a correlation close to 1 (not exactly 1, since the average does not completely remove all trial-to-trial variability). 
However, correlation to average does not measure how well a model accounts for stimulus-independent variability caused by behavioral fluctuations.

We calculate $\rho_{\textrm{ta}}$ in the same way as $\rho_{\textrm{st}}$, but we first average the responses and predictions per frame across all video repeats, where $\bar{r}_{i}$ is a response averaged over stimulus repeats for a fixed neuron:  
\begin{equation}
    \rho_{\textrm{ta}} = \textrm{corr}(\mathbf{r}_{\textrm{ta}}, \mathbf{o}_{\textrm{ta}}) = \frac{\sum_{i}(\bar{r}_{i} - \bar{r})(\bar{o}_{i} - \bar{o})}{\sqrt{\sum_{i}(\bar{r}_{i} - \bar{r})^2 \sum_{i}(\bar{o}_{i} - \bar{o})^2}},
\end{equation}


The initial 50 frames of predictions and neuronal responses were excluded from all metrics calculations. 
This allowed a ``burn-in'' period for models relying on history to achieve better performance.

\section{Baseline models}
\label{baseline}
\sensorium~2023 was accompanied by three model baselines, representing the state of the art in the field at the beginning of the competition:

\textbf{GRU baseline} is a dynamic model with a 2D CNN core and gated recurrent unit (GRU) inspired by earlier work \citep{sinz2018stimulus}, but with more recently developed Gaussian readouts~\citep{Lurz2020-ua}, which improves performance. 
Conceptually, the 2D core transforms each individual frame of the video stimulus into a feature space which is subsequently processed by a convolutional GRU across time.
The Gaussian readout then learns the spatial preference of each neuron in the visual space (``receptive field''), by learning the position at which a vector from the feature space is extracted by bilinear interpolation from the four surrounding feature map locations.
This latent vector is multiplied with a weight vector learned per neuron (``embedding'') and put through a rectifying nonlinearity to predict the activity of the neuron at each time step. 

\textbf{Factorized baseline} is a dynamic model~\citep{vystrcilova2024,hoefling2022chromatic} with a 3D factorized convolution core and Gaussian readouts. 
In contrast to the GRU baseline, where the 2D CNN core does not interact with the temporal component, the factorized core learns both spatial and temporal filters in each layer. 

\textbf{Ensembled baseline} is an ensembled version of the above factorized baseline over 14 models.
Ensembling is a well-known tool to improve the model performance in benchmark competitions \citep{allenzhu2023understanding}.
As we wanted to encourage participants to focus on novel architectures and training methods beyond simple ensembling, only entries outperforming the ensembled baseline were considered candidates for competition winners.

\textbf{Training.} All baseline models were trained with batch size 40 in the following way: 
For each of the 5 animals 8 video snippets consisting of 80 consecutive video frames starting at a random location within the video 
were passed and the gradient accumulated over all animals before performing optimizing step.
We used early stopping with a patience of 5.

\section{Results and Participation}
\label{results}

\begin{table}
\centering
\begin{tabular}{l|cc|cc}
 & \multicolumn{2}{c|}{\textbf{Main track}} & \multicolumn{2}{c}{\textbf{Bonus track}} \\
\textbf{Model} & single-trial $\rho_{st} \uparrow$ & average $\rho_{ta} \uparrow$ & single-trial $\rho_{st} \uparrow$ & average $\rho_{ta} \uparrow$ \\ \hline
DwiseNeuro & \textbf{0.291} & \textbf{0.542} & \textbf{0.221} & \textbf{0.429} \\ 
Dynamic-V1FM & 0.265 & 0.493 & 0.183 & 0.336 \\ 
ViV{\small 1}T & 0.243 & 0.457 & 0.178 & 0.333 \\ 
\hline
Ensemble baseline & 0.197 & 0.371 & 0.129 & 0.241 \\
Factorized baseline & 0.164 & 0.321 & 0.121 & 0.223 \\
GRU baseline & 0.106 & 0.207 & 0.059 & 0.106 \\
\end{tabular}
\caption{Model performance of competition winners and baselines on both tracks.}
\label{tab:results}
\end{table}
In the four-month submission period, out of 44 registered teams, 22 teams submitted a combined total of 163 models (main track: 22 teams, 134 submissions, bonus track: 5 teams and 29 submissions). 
The strong baseline models were surpassed in both tracks by 48\% and 70\%, respectively (\tabl{tab:results}). 
Notably, the winning model -- DwiseNeuro -- outperformed all other models on both tracks by a fairly decent margin, and the difference seemed even stronger on the out-of-domain data than on the main track. 
In contrast, the runner-up solution -- Dynamic-V1FM -- had somewhat of an edge over the third place -- ViV1T -- on the main track, but both were on par on the out-of-domain data (\tabl{tab:results}).
In the following we describe the three winning teams' approaches.
\begin{figure*}[t]
        \centering
        \includegraphics[trim=0 0 0 0, clip,width=\linewidth]{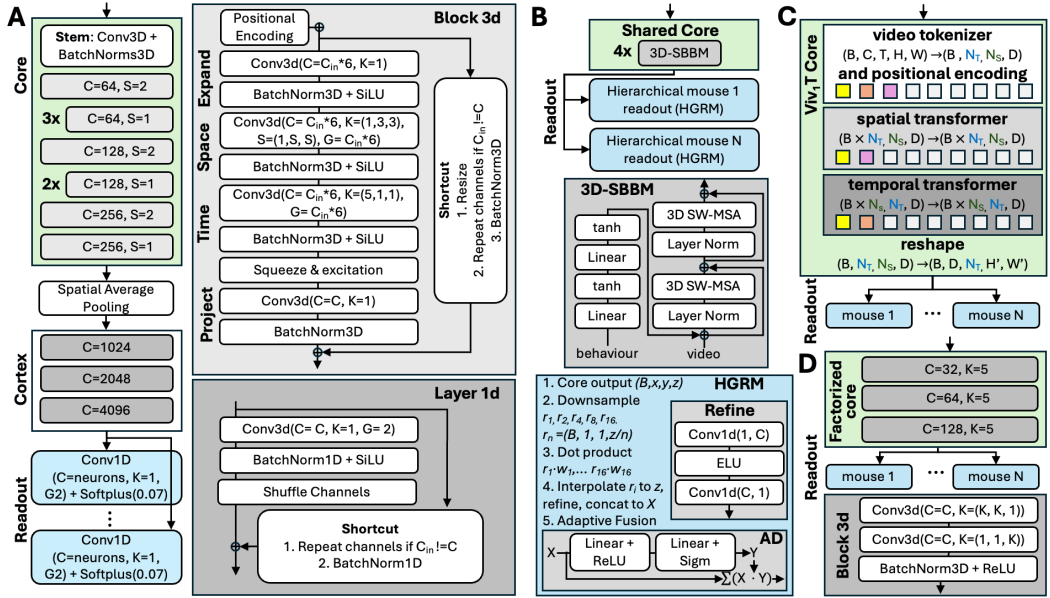}
        \caption{\textbf{Architectures of winning solutions.} 
        Across all subplots: \textit{C}: number of output channels in convolution layers, \textit{C}\textsubscript{\textit{in}}: number of input channels, \textit{K}: size of convolution kernels, \textit{S}: stride, \textit{G}: number of groups for convolution channels, \textit{B}: batch size.
        Core: green, readout: blue.
        \textbf{A:}~DwiseNeuro. 
        The core is based on 3D factorised convolutions. 
        The only solution whose readout was not based on the Gaussian readout \citep{Lurz2020-ua}.
        \textbf{B:}~Dynamic-V1FM. 
        The core is transformer-based, the Gaussian readout is extended to look in different resolution to the core output, then to fuse different resolutions. 
        Here $w$ represents the readout linear weights learnt for each neuron.
        \textbf{C:}~ViV{\small 1}T. The idea is to replace the core with a spatiotemporal transformer.
        \textbf{D:}~Ensembled factorized baseline.\vspace{-13pt}
        }
   \label{fig:winning_architectures}
\end{figure*}
\label{solutions}

\subsection{Place 1: DwiseNeuro}
\textbf{Architecture.}
DwiseNeuro consists of three main parts: core, cortex, and readouts. 
The core consumes sequences of video frames and mouse behavior activity in separate channels, processing temporal and spatial features. Produced features are aggregated with global average pooling over space. 
The cortex processes the pooled features independently for each timestep, increasing the channels. 
Finally, each readout predicts the activation of neurons for the corresponding mouse.

\textbf{Core.}
The first layer of the module is the stem. 
It is a point-wise 3D convolution for increasing the number of channels, followed by batch normalization. 
The rest of the core consists of factorised inverted residual blocks \citep{tan2019efficientnet, sandler2018mobilenetv2} with a \texttt{narrow -> wide -> narrow} channel structure (Fig. \ref{fig:winning_architectures}A). 
Each block uses (1)~absolute positional encoding \citep{vaswani2017attention} to compensate for spatial pooling after the core, (2)~factorized (1+1) convolutions \citep{tran2018closer}, (3)~parameter-free shortcut connections interpolating spatial sizes and repeating channels if needed, (4)~squeeze-and-excitation mechanism \citep{hu2018squeeze} to dynamically recalibrate channel-wise features, (5)~DropPath regularization \citep{larsson2016fractalnet, huang2016deep} that randomly drops the block's main path for each sample in the batch.
Batch normalization is applied after each layer.
SiLU activation \citep{elfwing2018sigmoid} is used after expansion and depth-wise convolutions. 

\textbf{Cortex.}
Spatial information accumulated through positional encoding was compressed by spatial global average pooling after the core, while the time dimension was unchanged.
The idea of the ``cortex'' is to smoothly increase the number of channels before the readout and there is no exchange of information across time.
First, the channels are split into two groups, then each group's channels are doubled as in a fully connected layer. 
Next, the channels are shuffled across the groups and concatenated. 
The implementation uses 1D convolution with two groups and kernel size one, with shuffling as in \citet{zhang2018shufflenet}. 
This procedure is repeated three times.
Batch normalization, SiLU activation, and shortcut connections with stochastic depth were applied similarly to the core.

\textbf{Readout.}
The readout is independent for each session, represented as a single 1D convolution with two groups and kernel size 1, 4096 input channels and the number of output channels equal to the number of neurons per mouse. 
It is followed by softplus activation as in \citet{hoefling2022chromatic}.

\textbf{Training.} The main changes compared to the baseline are introducing CutMix data augmentation \citep{yun2019cutmix}, removing normalization, and padding the frames to $64\times64$ pixels.
For more details on the training recipe, see Appendix~\ref{appendix:iromul-training}.


\textbf{Code.} Code is available at \url{https://github.com/lRomul/sensorium}

\subsection{Place 2: Dynamic-V1FM}


\textbf{Architecture.} Dynamic-V1FM (Dynamic V1 Functional Model), follows the pipeline proposed by \citet{wang2023towards}. It incorporates a \textit{shared core} module across mice and an \textit{unshared readout} module for individual mice.
The shared core module comprises four blocks of Layer Norms and 3D window based multi-head self-attention (MSA), inspired by the 3D swin transformer block \citep{liu2022video} combined with a two-layer behavioral multi-layer perceptron (MLP) module \citep{li2023vt}. The readout module is a Hierarchical Gaussian Readout Module (HGRM), which extends the Gaussian readout module \citep{Lurz2020-ua} by introducing a multi-layer design before the final linear readout (Fig. \ref{fig:winning_architectures}B).

\textbf{Ensemble Strategy.} As the readout module could support up to five levels of features and original layer is not downsampled and is always used as a base, four combinations of low-resolution features were traversed, resulting in  $C_{4}^{4}+C_{4}^{3}+C_{4}^{2}+C_{4}^{1}=1+4+6+4=15$ models, where $C_{n}^{k}$ is a binomial coefficient $C_{n}^{k}=\frac{n!}{k!(n-k)!}$ with $n$ elements and $k$ combinations.
Feature enhancement modules were also added to the low-resolution part of these 15 models, but the performance improvement was insignificant. 
As another set of 15 candidate models, they were included in the subsequent average ensemble strategy.
A model with the original Gaussian readout module was also trained as a baseline. 
The aforementioned 31 models were trained with a fixed random seed of 42, followed by an average ensemble of their predictions. 
For the final results of both competition tracks (the main track and the out-of-distribution track), the same model and ensemble strategy were used.

\textbf{Code.} Code is available at \url{https://github.com/zhuyu-cs/Dynamic-VFM}.

\subsection{Place 3: ViV{\small 1}T}
\textbf{Architecture}. 
The Vision Transformer (ViT, \citealt{dosovitskiy2021an}) was shown to be competitive in predicting mouse V1 responses to static stimuli~\citep{li2023vt}. 
Here, a factorized Transformer (ViV{\small 1}T) core architecture was proposed, based on the Video Vision Transformer by \citet{arnab2021vivit}, to learn a shared visual representation of dynamic stimuli across animals. 
The ViV{\small 1}T core contains three main components: (1)~a tokenizer that concatenates the video and behaviour variables over the channel dimensions and extracts overlapping tubelet patches along the temporal and spatial dimensions, followed by a factorized positional embedding which learns the spatiotemporal location of each patch; (2)~a spatial Transformer which receives the tubelet embeddings and learns the spatial relationship over the patches within each frame; (3)~a temporal Transformer receives the spatial embedding and learns a joint spatiotemporal representation of the video and behavioural information. 
This factorized approach allows to apply the self-attention mechanism over the space and time dimensions separately, reducing the size of the attention matrix and, hence, compute and memory costs.
Moreover, the vanilla multi-head attention mechanism is replaced by the FlashAttention-2~\citep{dao2023flashattention} and parallel attention~\citep{mesh-transformer-jax} to further improve model throughput. 

\textbf{Training}. 
The model was trained on all labeled data from the five original mice and the five competition mice. 
To isolate the performance differences solely due to the core architecture, the same shifter module, Gaussian readout~\citep{Lurz2020-ua}, data preprocessing and training procedure as the factorized baseline were employed. 
Finally, a Bayesian hyperparameter search~\citep{optuna_2019} of 20 iterations was performed to find an optimised setting for the core module (see \tabl{table:dunedin-hyperparameter}). 


\textbf{Ensemble}. The final submission was an average output of 5 models, initialized with different seeds. 

\textbf{Code.} Code is available at \url{https://github.com/bryanlimy/ViV1T}.

\section{Discussion}
\label{reflections}

\begin{table}
\centering
\begin{tabular}{c|c|c|c|c}
\textbf{Model} & \textbf{GPU} & \textbf{GPU memory} & \textbf{Batch Size} & \textbf{Wall Time} \\ \hline
DwiseNeuro & 2 $\times$ RTX A6000 & 48 Gb & 32 & 12h \\ 
Dynamic-V1FM & 8 $\times$ 2080Ti GPU & 11 Gb & 32 & 24h \\ 
ViV{\small 1}T & 1 $\times$ Nvidia A100 & 40 Gb & 60 & 20h \\ 
Factorized baseline & 1 $\times$ RTX A5000 & 24 Gb & 40 & 8h \\
GRU baseline & 1 $\times$ RTX A5000 & 24 Gb & 40 & 10h \\
\end{tabular}
\vspace{-5pt}
\caption{Training time for a single model (before ensembling).\vspace{-15pt}}
\label{train-time}
\end{table}

Different competition submissions explored different architectures. All winners employed architectures distinct from the baseline, but stayed roughly within the core-readout framework \citep{Antolik2016,Klindt2017}. 
Successful strategies included:
\begin{itemize}[leftmargin=*,nosep]
    \item Two out of three winning teams utilized transformer-based cores.
    \item Two teams also modified the readouts, but no team explicitly modeled temporal processing or interaction between neurons in the readout.
    \item However, the ``cortex'' module of the winning solution introduced several layers of nonlinear processing after spatial average pooling, effectively allowing all-to-all interactions.
    \item The winning solution kept the factorized 3D convolutions while introducing methods from computer vision models, such as skip connections and squeeze-and-excitation blocks. 
\end{itemize}
These observations suggest that classic performance-boosting methods from computer vision are also helpful to boost the performance for neural predictive models.
However, the impact of such architectural changes on the biologically meaningful insights, such as in \citet{Franke2022,burg2021learning, ustyuzhaninov2022}, still needs to be validated and requires additional research.

Another observation is that all three winning solutions included a mechanism for all-to-all interaction: the winning solution in the ``cortex'', the other two by using a transformer-based core. 
Thus, although the CNN has originally been modeled after primary visual cortex \citep{fukushima1980neocognitron}, it does not seem to provide the best inductive bias for modeling, at least mouse V1. 
Long-range interactions appear to be important. 
The current data does not allow us to resolve whether these long-range interactions actually represent visual information, as expected from lateral connections within V1 \citep{gilbert1983clustered,gilbert1989columnar}, or from more global signals related to the animal's behavior (which is also fed as input to the core). 
This will be an interesting avenue for future research.

Moving from static images to dynamic inputs in \sensorium~2023 increased the participation threshold markedly because of the higher demands on compute and memory. 
As a result, many models cannot be trained on freely available resources such as Colab or Kaggle anymore (\tabl{train-time}). 


\section{Conclusion}
\label{conclusion}
Predictive models are an important tool for neuroscience research and can deliver important insights to understand computation in the brain~\citep{Doerig2023-vq}.
We have seen that systematically benchmarking such models on shared datasets can boost their performance significantly. 
With the \sensorium\ benchmark we have successfully established such an endeavor for the mouse visual system. 
The 2023 edition successfully integrated lessons from 2022, such as including an ensemble to encourage participants to focus on new architectures. 
However, there are still ways to go to achieve a comprehensive benchmark for models of the visual system. 
Future iterations could include, among others, the following aspects:
\begin{itemize}[leftmargin=*,nosep]
    \item Use chromatic stimuli in the mouse vision spectrum \citep{hoefling2022chromatic, Franke2022}.
    \item Establish a benchmark for combining different data collection protocols \citep{azabou2024unified} or modalities \citep{antoniades2023neuroformer}.
    \item Focus not only on the predictive performance on natural scenes, but also on preserving biologically meaningful functional properties of neurons \citep{Walker2019-oq, ustyuzhaninov2022}.
    \item Extend beyond the primary visual cortex.
    \item Include more comprehensive measurements of behavioral variables.
    \item Include active behaviors of the animals.
\end{itemize}
We invite the research community to join us in this effort by continuing to participate in the benchmark and contribute to future editions.

\newpage
\bibliographystyle{apa-good}
\bibliography{reference}
\onecolumn
\newpage

\begin{ack}
The authors thank GWDG for the technical support and infrastructure provided. 
Computing time was made available on the high-performance computers HLRN-IV at GWDG at the NHR Center NHR@Göttingen.

FHS is supported by the Carl-Zeiss-Stiftung and acknowledges the support of the DFG Cluster of Excellence “Machine Learning – New Perspectives for Science”, EXC 2064/1, project number 390727645 as well as the German Federal Ministry of Education and Research (BMBF) via the Collaborative Research in Computational Neuroscience (CRCNS) (FKZ 01GQ2107). 
This work was supported by an AWS Machine Learning research award to FHS. 
MB and KW were supported by the International Max Planck Research School for Intelligent Systems. 
This project has received funding from the European Research Council (ERC) under the European Union’s Horizon Europe research and innovation programme (Grant agreement No. 101041669). 

The project received funding by the Deutsche Forschungsgemeinschaft (DFG, German Research Foundation) via Project-ID 454648639 (SFB 1528), Project-ID 432680300 (SFB 1456) and Project-ID 276693517 (SFB 1233).

This research was supported by National Institutes of Health (NIH) via National Eye Insitute (NEI) grant RO1-EY026927, NEI grant T32-EY002520, National Institute of Mental Health (NIMH) and National Institute of Neurological Disorders and Stroke (NINDS) grant U19-MH114830, NINDS grant U01-NS113294, and NIMH grants RF1-MH126883 and RF1-MH130416.  
This research was also supported by National Science Foundation (NSF) NeuroNex grant 1707400. The content is solely the responsibility of the authors and does not necessarily represent the official views of the NIH, NEI, NIMH, NINDS, or NSF.

This research was also supported by the Intelligence Advanced Research Projects Activity (IARPA) via Department of Interior/Interior Business Center (DoI/IBC) contract no. D16PC00003, and with funding from the Defense Advanced Research Projects Agency (DARPA), Contract No. N66001-19-C-4020. 
The US Government is authorized to reproduce and distribute reprints for governmental purposes notwithstanding any copyright annotation thereon. 
The views and conclusions contained herein are those of the authors and should not be interpreted as necessarily representing the official policies or endorsements, either expressed or implied, of IARPA, DoI/IBC, DARPA, or the US Government.

BML and WDW were supported by the United Kingdom Research and Innovation (grant EP/S02431X/1), UKRI Centre for Doctoral Training in Biomedical AI at the University of Edinburgh, School of Informatics.

\end{ack}







\appendix

\section{Winning solutions}

\subsection{Place 1 - DwiseNeuro}~\label{appendix:iromul-training}

\textbf{Analysis of Improvements.}
All of the score numbers are in for the main track during the competition live phase. An early model with depth-wise 3D convolution blocks achieved a score of $\approx$0.19. Implementing techniques from the core section, tuning hyperparameters, and training on ten mice instead of five data boosted the score to 0.25. Removing normalization improved the score to 0.27. The cortex and CutMix \citep{yun2019cutmix} increased the score to 0.276. Then, the $\beta$ value of Softplus was tuned, resulting in a score of 0.294. Lastly, adjusting drop rate and batch size parameters helped to achieve a score of 0.3. The ensemble of the basic and distillation \ref{appendix:iromul-training} training stages achieved a single-trial correlation of 0.2913. This is just slightly better than the basic training.

\begin{itemize}
    \item Learning rate warmup for the first three epochs from 0 to 2.4e-03
    \item cosine annealing last 18 epochs to 2.4e-05
    \item Batch size 32, one training epoch comprises 72000 samples
    \item Optimizer AdamW with weight decay 0.05
    \item Poisson loss
    \item Model EMA with decay 0.999
    \item CutMix with alpha 1.0 and usage probability 0.5
    \item The sampling of different mice in the batch is random by uniform distribution
\end{itemize}

\subsection{Place 2 - The Runner-up Solution \textit{Dynamic-V1FM}}

\begin{figure}[h]
  \centering
  \includegraphics[scale=0.4]{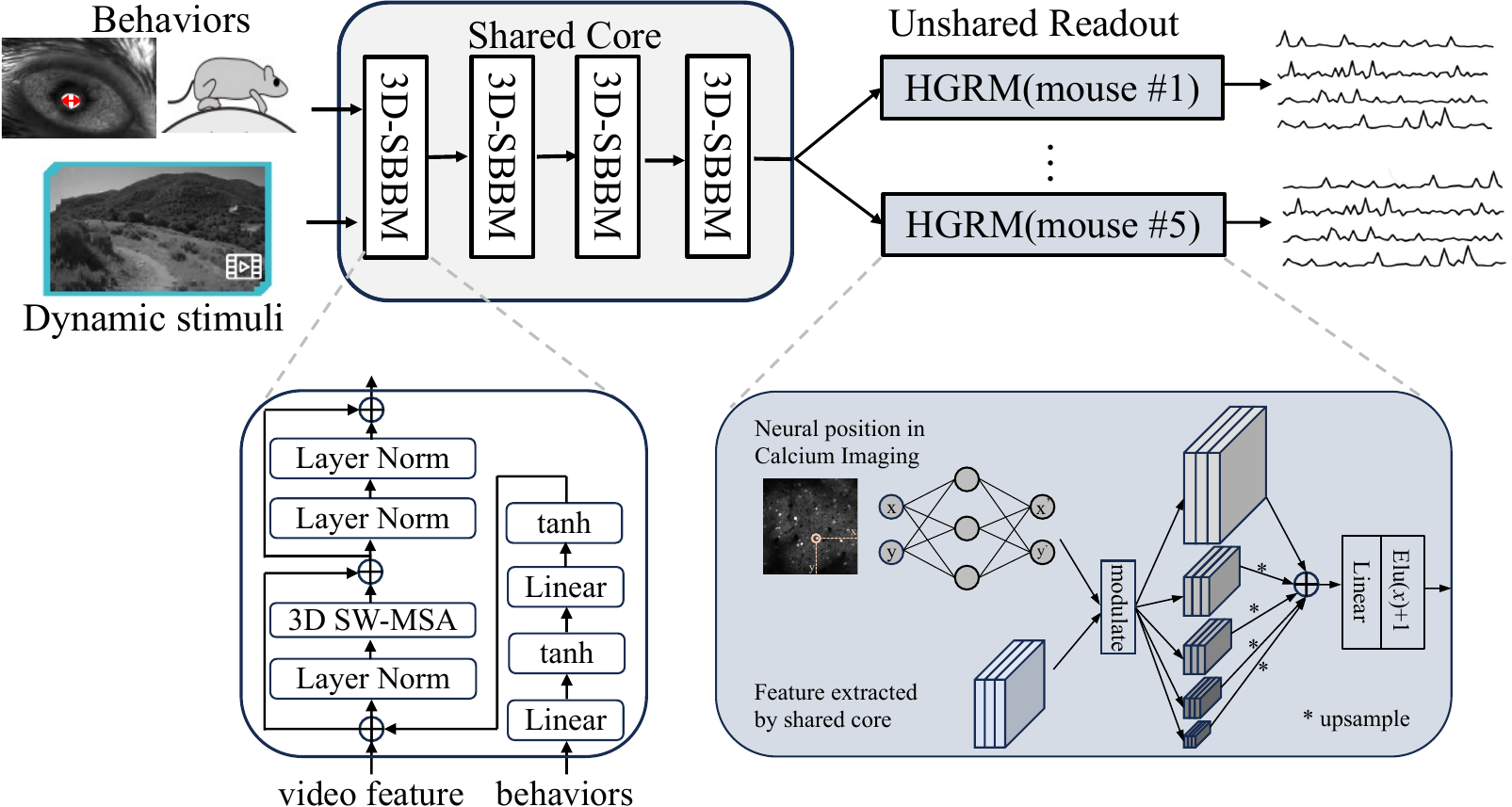}
  \caption{The overall architecture of Dynamic-V1FM. The core module consists of four 3D-SBBMs that process video and behavioral information, as detailed in the lower left. The unshared readout module includes five levels of features before linear readout in the lower right.}
  \label{V1FM}
\end{figure}

\textbf{Experiments}

\textbf{Training Details.} We trained Dynamic-V1FM using the training set of ten mice data provided by the competition, and tested it with only five mice data required for submission. Note that we did not employ any pre-training strategy and directly performed the evaluations required by the competition after training. During training, we used truncated normal initialization for the core module and the same initialization strategy for the readout module as \citet{Lurz2020-ua}. The whole model was optimized by AdamW optimizer \citep{loshchilov2017decoupled} with $(\beta_{1}, \beta_{2})=(0.9, 0.999)$, $\text{weight\_decay}=0.05$, and a batch size of 32. Each batch contained 30 frames of randomly sampled data. The peak learning rate was $1e^{-3}$, linearly warmed up with ratio $\frac{1}{3}$ for the first 600 iterations, then kept constant for the first 80 epochs and decreased to $1e^{-6}$ in the last 100 epochs with a cosine strategy. All the models used in the ensemble strategy shared the same training setting.

\textbf{Experimental Results.} On the live-test evaluation, the improvement of the core module, replacing 3D convolution with 3D swin transformer, resulted in an $R^2$ improvement of 0.045 (from 0.188 to 0.233). Enhancements in the readout module, replacing Gaussian readout to Hierarchical Gaussian readout, further improved the model by 0.018 (from 0.233 to 0.251). The final ensemble strategy yielded an overall prediction score of 0.276.

\textbf{Discussions}

We shall provide some thoughts on the V1FM design. Using combined data sets from multiple mice and a \textit{shared core} module for training is an efficient approach, although the subject-specific readout module strategy increases the difficulty of training the core module. This design could be viewed as a stronger regularization that may weaken the performance of the whole model. This problem might be alleviated by designing a new shared readout module that also relies on subject-specific information, such as mice identities and behavioral data. Specifically, we can use a readout module with dynamic weights \citep{chen2020dynamic} which is adjusted by mice identities and pupil size.

\subsection{Place 3 - ViV{\small 1}T}~\label{appendix:dunedin-hyperparameters}

\begin{table}[ht]
    \caption{ViV{\small 1}T core hyperparameter search space and their final settings. We performed Hyperband Bayesian optimization~\citep{li2017hyperband} with 20 iterations to find the setting that yield the best single trial correlation in the validation set. The resulting ViV{\small 1}T model contains 12M trainable parameters, about $13\%$ more than the factorized baseline.} \label{table:dunedin-hyperparameter}
    \begin{center}
    \begin{small}
    \begin{sc}
    \begin{tabular}{llr}
        \toprule
        hyperparameter & search space & final value \\
        \midrule
        Core & & \\
        embedding dim. & uniform, min: 8, max: 512, step: 8 & 112 \\
        learning rate & uniform, min: 0.0001, max: 0.01 & 0.0048 \\
        patch dropout & uniform, min: 0, max: 0.5 & 0.1338 \\
        drop Path & uniform, min: 0, max: 0.5 & 0.0505 \\
        pos. encoding & none, learnable, sinusoidal & learnable \\ 
        weight decay & uniform, min: 0, max: 1 & 0.1789 \\
        batch size & uniform, min: 1, max: 64 & 6 \\
        \midrule
        \multicolumn{3}{l}{Spatial Transformer} \\
        num. blocks & uniform, min:1, max: 8, step: 1 & 3 \\
        patch size & uniform, min: 3, max: 16, step: 1 & 7 \\
        patch stride & uniform, min: 1, max: patch size, step: 1 & 2 \\
        \midrule
        \multicolumn{3}{l}{Temporal Transformer} \\
        num. blocks & uniform, min:1, max: 8, step: 1 & 5 \\
        patch size & uniform, min: 1, max: 50, step: 1 & 25 \\
        patch stride & uniform, min: 1, max: patch size, step: 1 & 1 \\
        \midrule
        \multicolumn{3}{l}{multi-head attention (MHA) layer} \\
        num. heads & uniform, min: 1, max: 16, step: 1 & 11 \\
        head dim. & uniform, min: 8, max: 512, step: 8 & 48 \\
        MHA dropout & uniform, min: 0, max: 0.5 & 0.3580 \\ 
        \midrule
        \multicolumn{3}{l}{feedforward (FF) layer} \\
        FF dim. & uniform, min: 8, max: 512, step: 8 & 136 \\
        FF activation & Tanh, Sigmoid, ELU, GELU, SwiGLU & GELU \\
        FF dropout & uniform, min: 0, max: 0.5 & 0.0592 \\
        \bottomrule
    \end{tabular}
    \end{sc}
    \end{small}
    \end{center}
\end{table}
\subsection{Baseline architectures parameters}
\begin{table}
\centering
\begin{tabular}{|c|c|c|c|c|c|}
\hline
\textbf{Baseline} & \textbf{Core} & \textbf{Layers} & \textbf{Channels} & \textbf{Input Spatial} & \textbf{Spatial} \\
 &  & &  & \textbf{Kernels} & \textbf{ Kernels} \\\hline
GRU & Rotation-equivariant & 4  & 8 & $9\times9$ & $7\times7$ \\ \hline
3D Factorized & 3D factorized & 3 & 32, 64, 128 & $11\times11$ & $5\times5$ \\ \hline
\end{tabular}
\caption{\textbf{Core parameters for the baseline architectures}. Compared to the GRU baseline, the amount of channels in the core was increased sequentially.}
\label{base-params}
\end{table}

The \textbf{GRU baseline} used rotation-equivariant core from \cite{Ecker2018} with 8 rotations, resulting in 64 channels totally (8 channels $\times$ 8 rotations = 64). 
Inspired by \cite{Sinz2019} we used the GRU module after the core. It had 64 channels, and both input and recurrent kernels were $9 \times 9$.\\
For the \textbf{3D Factorized baseline}, we used the core inspired by \cite{hoefling2022chromatic, vystrcilova2024}. The temporal kernels were $11 \times 1$ in the 1st layer and $5 \times 1$ afterwards, same as the spatial ones (Tab. \ref{base-params}).\\
The Ensembled baseline cores were same as for the 3D Factorized baseline.
\subsection{Stability analysis}
\begin{table}[]
\centering
\begin{tabular}{l|cc}
 & \multicolumn{2}{c}{\textbf{Main track}}  \\
\textbf{Seed} & single-trial $\rho_{st} \uparrow$ & average $\rho_{ta} \uparrow$  \\ \hline
8	&0.1932&	0.3650\\
16	&0.1642&	0.3210\\
42	&0.1887&	0.3569\\
64	&0.1780&	0.3380\\
128	&0.1839&	0.3479\\
512	&0.1799&	0.3402\\
1024	&0.1865&	0.3528\\
2048	&0.1672&	0.3178\\
4096	&0.1734&	0.3305\\
16384	&0.1880&	0.3571\\
32768	&0.1933&	0.3661\\
131072	&0.1852&	0.3513\\
262144	&0.1839&	0.3488\\
1048576	&0.1943&	0.3674\\
\hline
mean	&0.1828&	0.3472\\
std	&0.0094&	0.0159\\
\end{tabular}
\caption{We used seeds 8, 16, 42, 64, 128, 512, 1024, 2048, 4096, 16384, 32768, 131072, 262144, 104857 to ensemble the factorized benchmark. 
Here we provide the individual performance of the models on the final test set to analyse how much performance depended on the seed.}
\label{tab:stability}
\end{table}
\clearpage
\subsection{Supplementary materials}
\subsubsection{Dataset documentation and intended uses}
Dataset documentation is available at \url{https://gin.g-node.org/pollytur/sensorium_2023_dataset} (dataset stucture) and in the whitepaper \citep{turishcheva2023dynamic} (data collection methodology).
Intended usage examples (loading of the data and models training) are available here: \url{https://github.com/ecker-lab/sensorium_2023/tree/main/notebooks}.
\subsubsection{URL for data download}
Five competition mice: \url{https://gin.g-node.org/pollytur/sensorium_2023_dataset}\\
Five mice with ood responses \url{https://gin.g-node.org/pollytur/sensorium_2023_data/src/798ba8ad041d8f0f0ce879af396d52c7238c2730}.
\subsubsection{Croissant url}
As the croissant library currently does not support the Video and List data types (\url{https://github.com/mlcommons/croissant/issues/690}), we generated the high-level meta file using kaggle interface:
\url{https://github.com/ecker-lab/sensorium_2023/blob/croissant_file/sensorium-2023-metadata.json}
\subsubsection{Author statement}
Author bear all responsibility in case of violation of rights. 
Both data and code are available under Creative Commons Attribution-NonCommercial-NoDerivatives 4.0 International License.
\subsubsection{Hosting, licensing, and maintenance plan}
Following \sensorium~2022, data is hosted at \url{https://gin.g-node.org}, which is a publicly available platform, where data can be downloaded both via GUI or command line. The code is hosted in a public repository via \url{https://github.com}.
The data does not need maintenance.
Both data and code are available under Creative Commons Attribution-NonCommercial-NoDerivatives 4.0 International License.
In case of any problems with the data hosting webpage, the authors have local copies of data and would re-release it.
\clearpage
\newpage
\end{document}